\documentclass[preprint,12pt,authoryear]{elsarticle}

\usepackage[letterpaper,margin=1in]{geometry}
\usepackage[utf8]{inputenc}
\usepackage[T1]{fontenc}
\usepackage{lmodern}
\usepackage{amsmath,amssymb,amsthm,mathtools}
\usepackage{mathrsfs}
\biboptions{authoryear,round}
\usepackage{url}
\usepackage{adjustbox}
\usepackage{enumitem}
\usepackage{array,booktabs,longtable}
\usepackage{microtype}
\usepackage{xcolor}
\usepackage{listings}
\usepackage[colorlinks=true,linkcolor=blue,citecolor=blue,urlcolor=blue]{hyperref}
\hypersetup{
  pdftitle={A Reduction Library for Polynomial-Base Harmonic Numbers},
  pdfauthor={Jayanta Phadikar}
}
\numberwithin{equation}{section}

\lstdefinestyle{wl}{
  basicstyle=\ttfamily\small,
  columns=fullflexible,
  breaklines=true,
  frame=single,
  backgroundcolor=\color{gray!6},
  keepspaces=true
}

\theoremstyle{plain}
\newtheorem{theorem}{Theorem}[section]

\theoremstyle{definition}
\newtheorem{definition}[theorem]{Definition}
\newtheorem{example}[theorem]{Example}
\theoremstyle{remark}
\newtheorem{remark}[theorem]{Remark}

\newcommand{\K}{\mathbb K}
\newcommand{\C}{\mathbb C}

\newcommand{\Z}{\mathbb Z}

\newcommand{\Pcal}{\mathcal P}
\newcommand{\Acal}{\mathcal A}
\newcommand{\Hcal}{\mathcal H}

\newcommand{\qbinom}[2]{\genfrac{[}{]}{0pt}{}{#1}{#2}_{q}}

\begin{document}

\begin{frontmatter}

\title{\texorpdfstring{\vspace*{1.0cm}}{}A Reduction Library for Polynomial-Base Harmonic Numbers}

\author[aff1]{Jayanta Phadikar\texorpdfstring{\corref{cor1}}{}}
\ead{jayantap@wolfram.com}

\address[aff1]{Wolfram Research}

\cortext[cor1]{Corresponding author.}

\begin{abstract}
We develop a finite reduction library for multiple polynomial-base harmonic numbers, strict colored nested sums in which the denominator letters at each summation level are univariate polynomials.  The affine and ordinary finite multiple harmonic numbers appear as lower-complexity subclasses, corresponding respectively to degree-one polynomial letters and to the ordinary letter $k$.  The main mechanisms include local normalization, rational single-level descent, Euclidean division, partial fractions, factorization of polynomial letters into affine letters, quadratic splitting, exact summation of empty and polynomial-numerator levels, affine shift and lattice reductions, staircase and complement transformations, repeated-level Newton reductions, weak-to-strict diagonal decompositions, and terminal ordinary harmonic-number reductions.  The accompanying Mathematica package provides a compact executable reduction library for polynomial-base, affine, and ordinary finite harmonic-number objects, with many checked examples recorded in a supplementary data-mine notebook.  The current supplementary rule inventory indexes roughly 670 reduction, guard, and normalization entries, of which about 160 are named family-level entries.  The library is intentionally conservative: rules are applied only under explicit hypotheses, such as absence of poles on the finite summation range, integer-power assumptions for partial-fraction descent, branch-safe scaling, finite factorization over an allowed coefficient extension, and, for telescoping, a verifiable certificate.

\end{abstract}


\end{frontmatter}

\section{Introduction}

Finite harmonic sums and their nested generalizations occur throughout symbolic summation, Mellin-space calculations, special-function expansions, and finite approximations to multiple zeta values.  The ordinary finite multiple harmonic numbers are strict chains built from the denominator letter $k$.  Colored versions add numerator factors $z^k$, and generalized harmonic sums, cyclotomic sums, $S$-sums, and $Z$-sums introduce larger alphabets of summand letters.  The algorithmic theory around these objects is well developed: Vermaseren's SUMMER algorithms handle nested symbolic sums involving harmonic series, binomial coefficients, and denominators \citep{Vermaseren1999}; Bl\"umlein and Kurth studied harmonic sums and Mellin transforms up to two-loop order \citep{BluemleinKurth1999}; Remiddi and Vermaseren introduced harmonic polylogarithms whose expansions and Mellin transforms are governed by harmonic sums \citep{RemiddiVermaseren2000}; and Moch, Uwer, and Weinzierl developed the $Z$-sum and $S$-sum formalism for expansions of transcendental functions and multi-loop integrals \citep{MochUwerWeinzierl2002}.

A second important line is the HarmonicSums and Sigma ecosystem.  Ablinger's HarmonicSums package and the works of Ablinger, Bl\"umlein, Raab, and Schneider treat harmonic sums, generalized harmonic sums, cyclotomic sums, binomially weighted sums, associated iterated integrals, structural relations, analytic continuation, and basis reductions \citep{Ablinger2010,AblingerBluemleinSchneider2011,AblingerBluemleinSchneider2013,AblingerBluemleinRaabSchneider2014,Ablinger2014}.  Schneider's difference-field and difference-ring methods, implemented in Sigma and related packages, provide a general symbolic summation framework, including telescoping and transformations of definite nested sums into indefinite nested product-sum expressions \citep{Karr1981,Schneider2013a,Schneider2013b,Schneider2014}.  These works form the natural background for any systematic treatment of finite nested sums.

The present paper focuses on a related but deliberately narrower problem.  We begin with strict finite nested sums whose denominator letters are univariate polynomial bases $P(k)$, possibly several at the same level and possibly multiplied by colors $z^k$.  We call these objects multiple polynomial-base harmonic numbers.  The affine case, where every letter is $ak+b$, is a degree-one subfamily; the ordinary multiple harmonic numbers, where every letter is $k$, are a further subfamily.  Thus we study one large container and a hierarchy of simpler target classes.  In the language of the monoidal alphabets framework of \citep{Phadikar2026}, many harmonic-sum constructions can be organized in ordinary, affine, and polynomial-base finite harmonic-number containers; the present paper studies explicit reductions inside and between those containers.  This is useful because harmonic-sum calculations often produce finite linear combinations of polynomial-base objects; reducing those objects can collapse complicated intermediate expressions to simple harmonic numbers, finite multiple harmonic numbers, or elementary endpoint terms.

Our purpose is not to list all identities among such sums.  Shuffle, stuffle, quasi-shuffle, duality, and symmetry identities are central in the theory, but many of them are relations among several objects rather than reductions of one object.  Here a rule is counted as a reduction only if it transforms one individual polynomial-base object, affine harmonic number, or ordinary harmonic number into a finite linear combination of simpler objects.  The simplification may be a class descent from polynomial to affine, a descent from affine to ordinary colored harmonic sums, a reduction of depth, a removal of polynomial numerator levels, or a terminal formula in classical harmonic numbers or elementary finite sums.

The contribution of the paper is therefore a conservative finite reduction calculus for a polynomial-base container.  It is complementary to $S$-sum, cyclotomic-sum, and difference-field frameworks: instead of providing a universal summation algorithm, it isolates explicit algebraic and finite-summation reductions that are easy to verify and useful as a data mine of polynomial-base collapses.

The paper is organized as follows.  Section \ref{sec:defs} defines ordinary, affine, and polynomial-base harmonic numbers and fixes the reduction hierarchy.  Section \ref{sec:mpbhn} gives the main polynomial-base reductions.  Section \ref{sec:affine} treats the affine case as the degree-one polynomial-base case.  Section \ref{sec:mhn} records the ordinary finite multiple harmonic target layer.  Section \ref{sec:pipelines} gives representative reduction cascades.  Section \ref{sec:implementation} describes the Mathematica implementation and the data-mine notebook.  The final section summarizes the scope and conclusions.

\section{Definitions and reduction hierarchy}\label{sec:defs}

Let $\K\subseteq\C$ be a coefficient field.  In reductions we allow finite extensions of $\K$ when polynomial letters are split into algebraic linear factors or when roots of unity are introduced.  The upper limit is a nonnegative integer $n$ in direct evaluations, although the formulas are stated symbolically in $n$ when no pole or endpoint ambiguity occurs.  We use descending strict chains
\[
        n\ge k_1>k_2>\cdots>k_r\ge1.
\]

\begin{definition}[Ordinary and colored finite multiple harmonic numbers]
For weights $s_1,\ldots,s_r$ and colors $z_1,\ldots,z_r$, define the colored finite multiple harmonic number
\begin{equation}\label{eq:colored-mhn}
H_n(s_1,\ldots,s_r;z_1,\ldots,z_r)
=
\sum_{n\ge k_1>\cdots>k_r\ge1}
\prod_{j=1}^r \frac{z_j^{k_j}}{k_j^{s_j}} .
\end{equation}
If all colors are $1$, we write
\begin{equation}\label{eq:mhn}
H_n(s_1,\ldots,s_r)=
\sum_{n\ge k_1>\cdots>k_r\ge1}
\prod_{j=1}^r k_j^{-s_j}.
\end{equation}
The empty depth object is $1$.
\end{definition}

\begin{definition}[Multiple affine harmonic numbers]
A level may contain several affine letters.  Let
\[
        L_{j,\nu}(k)=a_{j,\nu}+b_{j,\nu}k,
        \qquad b_{j,\nu}\ne0.
\]
The corresponding multiple affine harmonic number is
\begin{equation}\label{eq:mahn}
\Acal_n(\mathbf s,\mathbf L;\mathbf z)=
\sum_{n\ge k_1>\cdots>k_r\ge1}
\prod_{j=1}^r z_j^{k_j}
\prod_{\nu=1}^{m_j} L_{j,\nu}(k_j)^{-s_{j,\nu}} .
\end{equation}
We assume that positive denominator powers have no pole on the relevant finite index range.
\end{definition}

\begin{definition}[Multiple polynomial-base harmonic numbers]
Let $P_{j,\nu}(k)\in\K[k]$ be nonzero univariate polynomial letters.  The multiple polynomial-base harmonic number (polynomial-base harmonic number) is
\begin{equation}\label{eq:mpbhn}
\Pcal_n(\mathbf s,\mathbf P;\mathbf z)=
\sum_{n\ge k_1>\cdots>k_r\ge1}
\prod_{j=1}^r z_j^{k_j}
\prod_{\nu=1}^{m_j} P_{j,\nu}(k_j)^{-s_{j,\nu}} .
\end{equation}
A level with $m_j=0$ is called empty.  Nonpositive integer exponents turn denominator letters into polynomial numerator factors.
\end{definition}

The inclusions are immediate:
\begin{equation}\label{eq:hierarchy}
        \Hcal_{\rm ordinary}\subset \Hcal_{\rm colored}\subset \Acal\subset \Pcal.
\end{equation}
The class $\Acal$ is the degree-one subfamily of $\Pcal$, and the ordinary class is obtained from $\Acal$ by the special affine letter $k$.

\subsection{What counts as a reduction}

We use a qualitative complexity order.  The leading components are depth, number of active levels, total degree of non-linear polynomial letters, number of non-linear letters, number of affine shifts, total number of letters, and expression size.  A rule is promoted to a reduction only when it decreases one of these components without increasing an earlier component, or when it descends from a larger class to a smaller target class in \eqref{eq:hierarchy}.  This criterion separates reductions from general relations.

\begin{remark}
The package and notebook contain guard rules and normalization rules.  These are essential for reliable computation, but they are not all listed as theorems.  The main text emphasizes reductions that are class-changing, depth-changing, or terminal.
\end{remark}

\section{Reduction of polynomial-base harmonic numbers}\label{sec:mpbhn}

This section records the main polynomial-base reductions.  Elementary local normalizations are included first because they make the later formulas canonical.

\subsection{Elementary normalization}

For a polynomial letter $P(k)$ and exponents $s,t$,
\begin{align}
        P(k)^0&=1,\label{eq:zeroexp}\\
        P(k)^{-s}P(k)^{-t}&=P(k)^{-(s+t)},\label{eq:duplicate}\\
        (cP(k))^{-s}&=c^{-s}P(k)^{-s}\quad(c\in\K^\times),\label{eq:scaling}\\
        c(k)^{-s}&=c^{-s}\quad \text{if }c(k)=c\ne0.\label{eq:constant-letter}
\end{align}
The scale rule is used only when it is branch-safe, for example for integer exponents or known positive real scale factors.  A known zero color annihilates the strict sum provided the denominator side has no possible positive-integer pole.  All-empty chains reduce to binomial or Gaussian-binomial expressions:
\begin{align}
\sum_{1\le k_r<\cdots<k_1\le n}1&=\binom nr,\label{eq:empty-binom}\\
\sum_{1\le k_r<\cdots<k_1\le n}q^{k_1+\cdots+k_r}&=q^{r(r+1)/2}\qbinom nr\qquad(q\ne1).\label{eq:empty-qbinom}
\end{align}
The supplementary notebook contains many such elementary examples.  They are useful for normalization, but the substantial reductions below are the main contribution.

\subsection{Depth-one rational descent}

At depth one, a polynomial-base harmonic number is a finite sum of a colored rational function in one variable whenever the exponents are integers.  Let
\[
        R(k)=\prod_{\nu=1}^m P_\nu(k)^{-s_\nu}\in\K(k).
\]
After cancellation and Euclidean division,
\[
        R(k)=Q(k)+\frac{S(k)}{D(k)},\qquad \deg S<\deg D.
\]
If $D$ factors over a finite extension $L/\K$ as
\[
        D(k)=c\prod_{\alpha\in\Omega}(k-\alpha)^{m_\alpha},
\]
then partial fractions give

\begin{equation}\label{eq:rational-descent}
\begin{aligned}
\sum_{k=1}^n z^k R(k)
&=\sum_{m\ge0} q_m \sum_{k=1}^n z^k k^m  \\
&\quad+
\sum_{\alpha\in\Omega}\sum_{h=1}^{m_\alpha} C_{\alpha,h}
\sum_{k=1}^n \frac{z^k}{(k-\alpha)^h} .
\end{aligned}
\end{equation}
For $z=1$, the polynomial sums are Faulhaber sums,
\begin{equation}\label{eq:faulhaber}
\sum_{k=1}^n k^m=\frac{B_{m+1}(n+1)-B_{m+1}}{m+1},
\end{equation}
while for $z\ne1$ they are obtained from the finite geometric sum by $(z\partial_z)^m$.  The remaining terms are affine one-letter sums over $L$.

\begin{example}
For $P(k)=k^2+3k+2=(k+1)(k+2)$,
\[
 {1\over P(k)}={1\over k+1}-{1\over k+2},
\]
so
\[
\sum_{k=1}^n {z^k\over k^2+3k+2}
=
\sum_{k=1}^n {z^k\over k+1}
-
\sum_{k=1}^n {z^k\over k+2}.
\]
This is a polynomial-to-affine descent.
\end{example}

\subsection{Polynomial factorization to affine letters}

Suppose each positive-power polynomial letter in a level factors into linear terms over a field extension $L$,
\[
        P(k)=c\prod_{\alpha\in\Omega}(k-\alpha)^{e_\alpha}.
\]
For integer $s>0$,
\[
        P(k)^{-s}=c^{-s}\prod_{\alpha\in\Omega}(k-\alpha)^{-se_\alpha}.
\]
A level containing several such letters becomes a product of affine factors and therefore an affine level.  If the product has several distinct affine letters at the same level, partial fractions may further split it into a finite linear combination of single-letter affine levels in the depth-one case, or into multi-letter affine levels in nested cases.

Important special cases include:
\begin{enumerate}[label=(\alph*)]
\item power-polynomial collapse, $P(k)=c(ak+b)^d$;
\item rational-root families, $P(k)=c\prod_i(k+h_i)^{e_i}$;
\item rising and falling factorial denominators, such as $(k+a)_m$;
\item repeated linear factors, which produce repeated affine poles;
\item quadratic splitting, $ak^2+bk+c=a(k-\alpha)(k-\beta)$.
\end{enumerate}

\begin{example}[Rising factorial]
Since
\[
        {1\over k(k+1)}={1\over k}-{1\over k+1},
\]
we have
\[
        \sum_{k=1}^n {z^k\over k(k+1)}
        =H_n(1;z)-\sum_{k=1}^n {z^k\over k+1}.
\]
The second term is affine and can be shifted further when desired.
\end{example}

\subsection{Polynomial numerator-level deletion}

If every exponent at a level is a nonpositive integer and the color at that level is $1$, the level summand is a polynomial in its summation index.  Let
\[
        M(k)=\prod_{\nu=1}^m P_\nu(k)^{-s_\nu}\in\K[k].
\]
Choose a finite antidifference, or prefix polynomial, $F$ satisfying
\[
        F(t)-F(u)=\sum_{u<k\le t}M(k).
\]
For an interior level $k_{j-1}>k_j>k_{j+1}$, summing out $k_j$ gives
\begin{equation}\label{eq:poly-level-delete}
\sum_{k_{j+1}<k_j<k_{j-1}}M(k_j)
=F(k_{j-1}-1)-F(k_{j+1}).
\end{equation}
Thus the depth drops by one, at the cost of multiplying neighboring levels by polynomial boundary factors.  Those boundary factors are represented by negative-exponent polynomial letters and are then handled by the same normalization rules.

\begin{example}
With an empty upper wall and $M(k)=k^2$,
\[
        \sum_{1\le k<\ell} k^2={\ell(\ell-1)(2\ell-1)\over6}.
\]
Inside a nested chain this deletes the $k$-level and multiplies the adjacent $\ell$-level by the displayed polynomial.
\end{example}

\subsection{Empty levels and empty blocks}

An empty level contributes only its color.  If $u<k<v$,
\[
        \sum_{u<k<v}q^k=\begin{cases}
        v-u-1,&q=1,\\[0.25em]
        {q^{u+1}-q^v\over1-q},&q\ne1.
        \end{cases}
\]
Consequently an empty level can be deleted and replaced by boundary terms involving the neighboring indices.  Consecutive empty color-one blocks are counted by binomial interval coefficients:
\[
        \#\{u<j_m<\cdots<j_1<v\}=\binom{v-u-1}{m}.
\]
Common-color empty blocks give Gaussian-binomial interval factors.  These rules are genuine depth reductions.

\subsection{Staircase weak-chain transport}

Suppose the polynomial letters have staircase shifts: the $j$th level contains polynomials that simplify after the change
\[
        m_j=k_j+j-1.
\]
The strict inequalities $k_1>k_2>\cdots>k_r$ become weak inequalities
\[
        m_1\ge m_2\ge\cdots\ge m_r.
\]
The weak chain is then expanded into strict chains by merging consecutive equality blocks.  If a block $j,\ldots,j+\rho-1$ is merged, all polynomial letters in the block are placed at one level and the colors multiply.  Thus the weak object becomes a finite linear combination of strict polynomial-base harmonic numbers, often with simpler polynomial letters.

\begin{example}
A chain involving $k$, $k+1$, and $k+2$ can be transported to repeated ordinary letters after $m_j=k_j+j-1$.  The price is the weak-to-strict diagonal expansion, whose diagonal terms are lower-depth multi-letter levels.
\end{example}

\subsection{Complement reversal}

Some polynomial letters are naturally written in a reflected variable $n+h-k$.  If, for a fixed integer $h$, every transformed letter
\[
        P_{j,\nu}(n+h-y)
\]
is independent of $n$ as a polynomial in $y$, then the finite chain can be reversed by the substitution $y=n+h-k$.  The level order reverses, colors invert and contribute endpoint factors, and the resulting lower cutoff is removed by strict-interval subtraction.  The rule is accepted only when all transformed letters are polynomial in the new variable and no color inversion or pole condition is unsafe.

\subsection{Repeated-level Newton reduction}

If a block consists of repeated identical level factors
\[
        a_k=z^k\prod_{\nu=1}^mP_\nu(k)^{-s_\nu},
\]
then
\[
        \sum_{n\ge k_1>\cdots>k_r\ge1}a_{k_1}\cdots a_{k_r}
        =e_r(a_1,\ldots,a_n),
\]
where $e_r$ is an elementary symmetric polynomial.  Newton identities express $e_r$ in terms of power sums
\[
        p_m=\sum_{k=1}^n a_k^m.
\]
For instance,
\[
        e_2={p_1^2-p_2\over2},\qquad
        e_3={p_1^3-3p_1p_2+2p_3\over6}.
\]
Thus repeated depth can be collapsed to products and sums of depth-one polynomial-base harmonic numbers.

\subsection{Certificate telescoping}

A depth-one or local level summand $T(k)$ can be deleted if a certificate $G(k)$ is available and verified:
\[
        T(k)=G(k+1)-G(k).
\]
Then
\[
        \sum_{u<k<v}T(k)=G(v)-G(u+1).
\]
This is a powerful conditional reduction.  The present rule base treats it conservatively: a proposed certificate must be checked algebraically, and no universal telescoping search is claimed.

\section{Affine-case reductions}\label{sec:affine}

Affine harmonic numbers are degree-one polynomial-base harmonic numbers.  They are kept as a separate notation because many polynomial reductions naturally stop at affine letters before descending further.

\subsection{Affine normalization and partial fractions}

At one level,
\[
        \prod_{\nu=1}^m(a_\nu+b_\nu k)^{-s_\nu}
\]
can be simplified by extracting branch-safe scale factors, combining proportional letters, and applying partial fractions when integer powers and distinct affine poles permit.  For example,
\[
        {1\over(k+a)^p(k+b)^q}
        =\sum_{i=1}^p {A_i\over(k+a)^i}
        +\sum_{j=1}^q {B_j\over(k+b)^j},\qquad a\ne b.
\]
This reduces the number of active affine letters at the level.

\subsection{Uniform integer shift}

If every level has the same shift $k+h$, then the substitution $m=k+h$ preserves strict order:
\[
        n\ge k_1>\cdots>k_r\ge1
        \quad\Longleftrightarrow\quad
        n+h\ge m_1>\cdots>m_r\ge h+1.
\]
The interval $h<m\le n+h$ is written as an upper ordinary chain minus lower endpoint chains.  At depth one,
\begin{equation}\label{eq:shift-depth-one}
\sum_{k=1}^n {z^k\over(k+h)^s}
=z^{-h}\left(H_{n+h}(s;z)-H_h(s;z)\right),
\end{equation}
with the obvious uncolored specialization.

\subsection{Depth-two shifted-boundary reductions}

The implementation includes one-sided depth-two shift reductions.  A lower shifted letter such as $k_2+h$ is transported by moving the lower index and subtracting the finite strip created at the boundary.  An upper shifted letter has the analogous correction with the upper limit extended to $n+h$ and a finite family of diagonal bands subtracted.  These rules are useful when a factorized polynomial produces an affine shift at exactly one level.

\subsection{Staircase and complement affine cases}

The affine staircase rule is the degree-one version of the polynomial staircase rule.  If the shifts are compatible with
\[
        k_j+h_0+j-1,
\]
then the strict chain becomes a weak ordinary chain, which is expanded into strict blocks.  Complement reversal likewise sends affine letters in $n+h-k$ to ordinary affine letters in the reversed index, with colors inverted and endpoint intervals subtracted.

\subsection{Affine lattice and root-of-unity filtering}

For a depth-one affine letter $ak+b$ with $a\in\Z_{>1}$, $b\in\Z_{\ge0}$, unit color, positive integer weight, and no finite pole, the congruence condition $j\equiv b\pmod a$ may be enforced by a root-of-unity filter:
\[
        {\bf 1}_{j\equiv b\,({\rm mod}\,a)}
        ={1\over a}\sum_{\ell=0}^{a-1}\zeta_a^{\ell(j-b)}.
\]
This gives
\begin{equation}\label{eq:root-unity-affine}
\sum_{k=1}^n{1\over(ak+b)^s}
={1\over a}\sum_{\ell=0}^{a-1}\zeta_a^{-b\ell}
\left(H_{an+b}(s;\zeta_a^\ell)-H_b(s;\zeta_a^\ell)\right).
\end{equation}
This is implemented only for a guarded depth-one case.  The broader residue-class theory is natural, but the package deliberately avoids unsafe automatic residue decompositions.

\section{Ordinary finite multiple harmonic target reductions}\label{sec:mhn}

The ordinary layer is the terminal class.  The  package includes a copied ordinary special-rule layer for selected symbolic reductions and uses the built-in finite multiple harmonic-number object for numerical upper limits.

\subsection{Zero, unit, and negative-unit weights}

The all-zero ordinary chain is elementary:
\[
        H_n(\underbrace{0,\ldots,0}_{r})=\binom nr.
\]
The all-one chain can be expressed using complete Bell polynomials:
\begin{equation}\label{eq:all-one-bell}
H_n(\underbrace{1,\ldots,1}_{r})
=\frac{1}{r!}B_r\left(x_1,\ldots,x_r\right),
\qquad
x_m=(-1)^{m-1}(m-1)!H_n(m).
\end{equation}
Similarly, all negative-one chains are elementary symmetric polynomials in $1,2,\ldots,n$ and can be computed from Newton identities using the power sums $\sum_{k=1}^n k^m$.

\subsection{Depth-two and depth-three terminal tables}

The ordinary special layer contains explicit low-depth formulas for several weights in $\{-1,0,1\}$.  These include depth-two cases such as $(-1,-1)$, $(-1,0)$, $(0,1)$, and $(1,1)$, and selected depth-three cases.  The package also includes rules for zero-weight positions with colors and for repeated ordinary levels.  In the final paper version these formulas can be inserted as a table after the complete ordinary finite multiple harmonic rule set is finalized.

\subsection{Repeated ordinary levels}

Repeated ordinary or colored levels are reduced by the same Newton mechanism as in the polynomial-base case.  If
\[
        a_k={z^k\over k^s},
\]
then
\[
        H_n(\underbrace{s,\ldots,s}_{r};\underbrace{z,\ldots,z}_{r})
        =e_r(a_1,\ldots,a_n),
\]
which is expressed in terms of depth-one sums
\[
        \sum_{k=1}^n {z^{mk}\over k^{ms}}=H_n(ms;z^m).
\]
This is a genuine depth reduction of a repeated block.

\section{Representative reductions, implementation, and evaluations}\label{sec:pipelines}

This section combines representative mathematical reductions with the corresponding Mathematica evaluations from the data-mine notebook.  The examples are deliberately short: each formula illustrates one rule family and each code block shows the corresponding special-rule call.

\subsection{Representative reduction pipelines}

\begin{example}[Polynomial to affine to ordinary]
For $P(k)=k^2+3k+2$,
\[
\sum_{k=1}^n{z^k\over P(k)}
=\sum_{k=1}^n{z^k\over k+1}-\sum_{k=1}^n{z^k\over k+2}.
\]
When $z=1$, both affine sums reduce to shifted harmonic numbers.  For general color, the current package avoids reducing to special functions such as Lerch transcendent or polylogarithms and instead returns only the supported finite harmonic targets.
\end{example}

\begin{example}[Power-polynomial collapse]
If $P(k)=c(ak+b)^d$, then
\[
        P(k)^{-s}=c^{-s}(ak+b)^{-ds}.
\]
Thus the polynomial-base letter is actually affine with weight multiplied by $d$.  If $a=1$ and $b$ is a nonnegative integer, the shifted affine rule descends further to ordinary finite harmonic sums.
\end{example}

\begin{example}[Polynomial numerator deletion]
For an interior polynomial numerator level, finite power sums delete the level.  For instance,
\[
        \sum_{u<k<v}(k^2+1)
        ={(v-1)v(2v-1)-u(u+1)(2u+1)\over6}+v-u-1.
\]
The deleted level is replaced by polynomial boundary multipliers on neighboring levels.
\end{example}

\begin{example}[Repeated quadratic levels]
Let
\[
        a_k={z^k\over(k^2+1)^s}.
\]
Then
\[
        \sum_{n\ge i>j\ge1}a_i a_j
        ={1\over2}\left(\left(\sum_{k=1}^n a_k\right)^2-
        \sum_{k=1}^n a_k^2\right).
\]
This reduces depth two to depth-one polynomial-base sums.
\end{example}

\subsection{Mathematica implementation}\label{sec:implementation}

The supplementary implementation is the package
\[
        \texttt{PolynomialBaseHarmonicReduce.wl}.
\]
It can be loaded from the directory of the companion notebook by
\begin{center}
\fbox{\parbox{0.92\linewidth}{\ttfamily\small
Get[FileNameJoin[\{NotebookDirectory[], "PolynomialBaseHarmonicReduce.wl"\}]];
}}
\end{center}
It contains polynomial-base, affine, and ordinary special reductions in one package.  The public heads and rule-entry functions are:
\begin{description}[leftmargin=0.4in,style=nextline]
\item[\texttt{MultiplePolybaseHarmonicNumber}] Formal and evaluable head for polynomial-base harmonic numbers.  It takes an upper limit, weights, polynomial letters, and optionally colors.
\item[\texttt{MultipleAffineHarmonicNumber}] Formal and evaluable head for the affine degree-one subclass.  A letter \texttt{\{a,b\}} denotes $a+bk$.
\item[\texttt{MPBHNSpecialRules}] Applies the polynomial-base symbolic reduction layer and returns the original object if no supported rule applies.
\item[\texttt{MAHNSpecialRules}] Applies the affine symbolic reduction layer and returns the original affine object if no supported rule applies.
\item[\texttt{MHNSpecialRules}] Applies the ordinary finite multiple harmonic-number special-rule layer, with fallback to the ordinary finite harmonic-number object.
\end{description}

Polynomial letters are represented by coefficient lists: \texttt{\{c0,c1,...,cm\}} denotes $c_0+c_1k+\cdots+c_mk^m$.  Weights may be scalars, vectors with one letter per level, or nested lists allowing several letters at one level.  Colors are optional and default to $1$ at every level.  The direct object \texttt{MultiplePolybaseHarmonicNumber} evaluates by the finite recursive definition when the upper limit is a nonnegative integer and the input is parameter-free enough.  For symbolic upper limits it attempts the same conservative special-rule layer used by \texttt{MPBHNSpecialRules}; if no rule is applicable, the expression is left unevaluated rather than guessed.

The package implements four main layers: normalization and guards, polynomial-base special reductions, affine reductions, and ordinary terminal reductions. The current supplementary rule inventory indexes roughly 670 reduction, guard, and normalization entries, of which about 160 are named family-level entries; the main text concentrates on the substantial mechanisms rather than listing every elementary case.  The normalization layer includes canonical input shapes, zero-exponent deletion, constant extraction, duplicate/proportional letter fusion, pole detection, safe zero-color annihilation, and small-upper-limit expansion.  The polynomial-base layer includes monomial collapse, factorization to affine letters, quadratic splitting, rational depth-one partial fractions, polynomial numerator-level deletion, empty-block deletion, repeated-level Newton reductions, staircase transport, complement reversal, and conservative no-rule return.  The affine layer includes ordinary-letter recognition, affine partial fractions, uniform shift windows, shifted-boundary depth-two reductions, staircase transport, complement reversal, guarded depth-one root-of-unity lattice reduction, and repeated affine levels.  The ordinary layer includes all-zero chains, all-one chains through Bell polynomials, all-negative-one chains through power-sum Newton identities, selected low-depth tables, repeated ordinary blocks, zero-weight rules, and colored terminal cases.

The supplementary data-mine notebook contains examples that are too small or too implementation-specific for the main text, including elementary normalizations, guard tests, direct finite checks, partial reductions, and rule-family inventories.  Some inventory items are guards or specified families rather than standalone theorem-level reductions; the paper therefore highlights the substantial implemented mechanisms and uses the data mine as the complete experimental catalogue.

\subsection{Ten compact data-mine evaluations}

We write $G_s(x)$ for the generalized harmonic number \texttt{HarmonicNumber[x,s]} and $G_1(x)=\texttt{HarmonicNumber[x]}$.  The notation $H_N(s_1,\ldots,s_r;z_1,\ldots,z_r)$ is the colored finite multiple harmonic number of \eqref{eq:colored-mhn}; colors are omitted when all are $1$.

\paragraph{E1. Depth-one rational partial fractions}
\[
\sum_{k=1}^n {1\over k(k+1)}=1+H_n(1)-H_{n+1}(1).
\]
\begin{lstlisting}[style=wl]
MPBHNSpecialRules[n, {{1, 1}}, {{{0, 1}, {1, 1}}}, {1}]
(* 1 + HarmonicNumber[n] - HarmonicNumber[1+n] *)
\end{lstlisting}

\paragraph{E2. Repeated linear pole expansion}
\[
\sum_{k=1}^n {1\over (k+1)^3}=H_{n+1}(3)-1.
\]
\begin{lstlisting}[style=wl]
MPBHNSpecialRules[n, {{2, 1}}, {{{1, 1}, {1, 1}}}, {1}]
(* -1 + HarmonicNumber[1+n, 3] *)
\end{lstlisting}

\paragraph{E3. Improper rational quotient}
\[
\sum_{k=1}^n {k\over (k+1)^2}=H_{n+1}(1)-H_{n+1}(2).
\]
\begin{lstlisting}[style=wl]
MPBHNSpecialRules[n, {{-1, 2}}, {{{0, 1}, {1, 1}}}, {1}]
(* HarmonicNumber[1+n] - HarmonicNumber[1+n, 2] *)
\end{lstlisting}

\paragraph{E4. Shifted irreducible quadratic descent}
\[
\sum_{k=1}^n {1\over k^2+2k+2}
=-{i\over2}\bigl(G_1(n+1-i)-G_1(1-i)\bigr)
+{i\over2}\bigl(G_1(n+1+i)-G_1(1+i)\bigr).
\]
\begin{lstlisting}[style=wl]
MPBHNSpecialRules[n, {1}, {{2, 2, 1}}, {1}]
(* -(I/2) (-HarmonicNumber[1 - I] + HarmonicNumber[1 - I + n])
   + (I/2) (-HarmonicNumber[1 + I] + HarmonicNumber[1 + I + n]) *)
\end{lstlisting}

\paragraph{E5. Uniform affine shift window}
\[
\sum_{n\ge k_1>k_2\ge1}{1\over (k_1+1)^2(k_2+1)^3}
=1-H_{n+1}(2)+H_{n+1}(2,3).
\]
\begin{lstlisting}[style=wl]
MAHNSpecialRules[n, {{2}, {3}}, {{{1, 1}}, {{1, 1}}}, {1, 1}]
(* 1 - HarmonicNumber[1+n, 2] + MultipleHarmonicNumber[1+n,{2,3}] *)
\end{lstlisting}

\paragraph{E6. Scaled uniform affine shift window}
\[
\sum_{n\ge k_1>k_2\ge1}{1\over (2k_1+2)^2(2k_2+2)^3}
={1\over32}\bigl(1-H_{n+1}(2)+H_{n+1}(2,3)\bigr).
\]
\begin{lstlisting}[style=wl]
MAHNSpecialRules[n, {{2}, {3}}, {{{2, 2}}, {{2, 2}}}, {1, 1}]
(* 1/32 (1 - HarmonicNumber[1+n, 2]
   + MultipleHarmonicNumber[1+n, {2, 3}]) *)
\end{lstlisting}

\paragraph{E7. Colored uniform affine shift window}
\[
\sum_{n\ge k_1>k_2\ge1}
{(1/2)^{k_1}(1/3)^{k_2}\over (k_1+1)^2(k_2+1)^3}
=6\left({1\over6}-{1\over3}H_{n+1}(2;1/2)
+H_{n+1}(2,3;1/2,1/3)\right).
\]
\begin{lstlisting}[style=wl]
MAHNSpecialRules[n, {{2}, {3}}, {{{1, 1}}, {{1, 1}}}, {1/2, 1/3}]
(* 6 (1/6 - 1/3 HarmonicNumber[1+n, 2, 1/2]
   + MultipleHarmonicNumber[1+n, {2, 3}, {1/2, 1/3}]) *)
\end{lstlisting}

\paragraph{E8. All-one ordinary chain}
\[
H_n(1,1)={1\over2}\left(H_n(1)^2-H_n(2)\right).
\]
\begin{lstlisting}[style=wl]
MHNSpecialRules[n, {1, 1}]
(* 1/2 (HarmonicNumber[n]^2 - HarmonicNumber[n,2]) *)
\end{lstlisting}

\paragraph{E9. All-minus-one ordinary chain}
\[
H_n(-1,-1)=-{n\over12}-{n^2\over8}+{n^3\over12}+{n^4\over8}.
\]
\begin{lstlisting}[style=wl]
MHNSpecialRules[n, {-1, -1}]
(* -n/12 - n^2/8 + n^3/12 + n^4/8 *)
\end{lstlisting}

\paragraph{E10. Colored repeated ordinary chain}
\[
H_n(2,2;1/2,1/2)
={1\over2}\left(H_n(2;1/2)^2-H_n(4;1/4)\right).
\]
\begin{lstlisting}[style=wl]
MHNSpecialRules[n, {2, 2}, {1/2, 1/2}]
(* 1/2 (HarmonicNumber[n,2,1/2]^2 - HarmonicNumber[n,4,1/4]) *)
\end{lstlisting}

Each of these notebook entries is accompanied there by a direct finite check comparing the special-rule output with the defining finite nested sum.

\section{Conclusions}

We introduced a finite reduction library for multiple polynomial-base harmonic numbers.  The central idea is to treat polynomial-base sums as the largest object class, affine sums as the degree-one intermediate target, and ordinary colored or uncolored multiple harmonic numbers as terminal targets.  The substantial rules reduce polynomial letters to affine letters by factorization and partial fractions, delete empty or polynomial-numerator levels by finite summation, collapse repeated blocks through Newton identities, transport staircase and complement geometries, and perform guarded affine descents to ordinary harmonic sums.  A Mathematica package and data-mine notebook accompany the theory and demonstrate many additional examples, including elementary normalization rules and safe-failure cases.  The calculus is not a universal summation algorithm: it is limited to explicit, checkable reductions under stated hypotheses, with conservative fallback when a rule would require unsafe factorization, branch choices, pole handling, or unverified telescoping.

Future work may extend the residue-class affine theory to higher depth, integrate stronger certificate search for telescoping, connect more directly with difference-field summation methods, and incorporate ordinary harmonic-number reduction tables more systematically.

\section*{Supplementary material}
The Mathematica package \texttt{PolynomialBaseHarmonicReduce.wl} and the accompanying data-mine notebook are intended as supplementary material.  The notebook contains additional evaluated examples, direct finite checks, elementary normalizations, and guard cases that are not expanded in the main text.

\section*{Declaration of competing interest}
The author declares no competing financial interests or personal relationships that could have appeared to influence the work reported in this paper.

\paragraph{AI-assisted preparation disclosure}
AI-assisted tools were used in a limited way to improve the clarity and presentation of selected parts of the manuscript. The mathematical ideas, definitions, results, proofs, computations, and overall substance are the author's own. All AI-assisted suggestions were carefully reviewed, and the author remains fully responsible for the accuracy, originality, and final form of the article.

\end{document}